\documentclass[a4paper]{article}
\usepackage{amsmath}
\usepackage{amssymb}
\usepackage{theorem}
\usepackage{latexsym}
\usepackage{subfig}
\usepackage{subfig}
\usepackage{graphicx,color}
\usepackage{multirow}
\usepackage{colortbl}
\usepackage{rotating}
\usepackage[table]{xcolor}
\usepackage{hhline}
\usepackage{floatrow}
\usepackage{mathrsfs}

\usepackage{todonotes}

\DeclareMathAlphabet{\mathantt}{OT1}{antt}{li}{it}
\DeclareMathAlphabet{\mathpzc}{OT1}{pzc}{m}{it}

\definecolor{lightgray}{gray}{0.95}
\newfloatcommand{capbtabbox}{table}[][\FBwidth]

\usepackage{float}
\floatstyle{plaintop}
\restylefloat{table}

\input amssym.def
\input amssym.tex

\def\der{{\rm d}}
\def\qed{\raise1pt\hbox{\vrule height5pt width5pt depth0pt}}
\title{An automatic dynamic balancer in a rotating mechanism with time-varying angular velocity}
\author{James A. Wright{\small$^1$}, Linyu Peng{\small$^{2}$}\footnote{Corresponding author. E-mail address: l.peng@aoni.waseda.jp} \\ \\
{\small 1. Department of Mathematics, University of Surrey, Guildford GU2 7XH, UK} \\
{\small 2.  Waseda Institute for Advanced Study,   Waseda University, Tokyo 169-8050, Japan}}
\date{}
\begin{document}
\maketitle
\begin{abstract}
We consider the system of a two ball automatic dynamic balancer attached to a rotating disc with nonconstant angular velocity. We directly compare the scenario of constant angular velocity with that when the acceleration of the rotor is taken into consideration. In doing so we show that there are cases where one must take the acceleration phase into consideration to obtain an accurate picture of the dynamics. Similarly we identify cases where the acceleration phase of the disc may be ignored. Finally, we briefly consider nonmonotonic variations of the angular velocity, with a view of maximising the basin of attraction of the desired solution, corresponding to damped vibrations.

{\bf Keywords:} attractor; automatic dynamic balancer; basin of attraction; mechanical system; ramped velocity
\end{abstract}
\section{Introduction}
There is a vast array of mechanical systems in which imbalance causes undesirable vibrations, e.g. washing machines, drills, vacuum cleaners, jet engines, fans, etc. In many instances the imbalance is caused by imperfections in the manufacturing process, however there are also cases where imbalance is caused by the operation of the machinery or resulting from sustained damage. One well known example in household appliances is
washing machines. When a washing machine which is filled with clothes spins, a large amount of imbalance is created due to the uneven distribution of the clothes. This causes the washing machine to vibrate. To reduce these vibrations, washing machines have a heavy concrete weight in the base. This makes them difficult to move, and still does not completely quell the vibrations.
Another example is the danger posed to aircraft by birds. If a bird is sucked through a jet engine of an aircraft it can damage the blades of the engine; in most cases leading to imbalance. So as to avoid dangerous vibrations and maintain control of the aircraft, the pilot is forced to shut down the engine. Even during normal operation, the vibrations resulting from jet-engines may cause weakness in the structure of aircraft.

It is clear that the unbalance of rotating parts in machinery is a common cause of undesirable vibrations and is a widespread problem. Such vibrations can cause damage to both the user and the machine itself, as well as create undesired noise pollution. For rotors with a fixed amount of imbalance, it is sufficient to balance the system only once. However, effects such as thermal deformation, material erosion and those mentioned above can cause the mass distribution to change, as a result the balancing procedure may have to be repeated. Furthermore, in cases where imbalance is caused as a result of the operation of machinery, it may not be possible to balance the system in advance.

An automatic dynamic balancer\footnote{Also referred to as an automatic ball balancer (ABB).} (ADB), is a passively controlled device which requires no external forces to eliminate the imbalance of rotating mechanisms. An ADB works using two or more weighted balls housed in a race which is filled with a viscous fluid. As the rotor spins the forces on the balls cause them to move so as to eliminate the imbalance in the rotor and hence quell the vibrations. The ADB system has already been extensively investigated in a large amount amount of literature by method of analysis, numerical simulation and even practical implementation, see for example \cite{Green1,Green2,Green3,Huang,Lee2} and the references contained therein. It was shown in \cite{Lee1,Lee2} that provided the angular velocity of the system, $\omega$, is greater than the critical (also called natural) angular velocity, $\omega_c$, an ADB may be successfully utilised to dampen the vibrations caused by imbalance.

However, while an ADB can reduce imbalance and minimise vibrations in a system, previous research has shown that it may also cause periodic motion far worse than the original vibrations \cite{Chung1}, which pose a greater risk of damage to both the user and the machine. Furthermore in \cite{Green1} it was shown that large transient time-spans are necessary in order to maximise the probability of achieving balance with an ADB\footnote{We refer here to the probability of achieving balance with random placement of the balls in the race, i.e. the relative size of the balanced state's basin of attraction in phase space $(\phi_1,\dots,\phi_n)$, where $\phi_i$ specifies the location of the $i$-th ball.}. Naturally, large transients are undesired, as damage may already be caused before balance is attained. For these reasons, ADB devices are not widely implemented in industry. In an attempt to address the prior mentioned problems there has also been research into modified ADB devices, see for example \cite{TKim,Rezaee1,Rezaee2,Sung}.

In previous literature it is assumed that the system rotates with constant
angular velocity, $\omega$, see for example \cite{Chung2,Ehyaei,Green2,Huang,Yang}. This is justified by assuming that the balls are held in position using a clamping mechanism while the rotor accelerates then released once the desired angular velocity is attained. The use of a clamping mechanism was first proposed in \cite{Ernst,Thearle} and has since become a standard approach when studying ADB systems.
This model ignores the effects of any vibrations which occur during the spin-up of the rotor, while the balls are clamped in position.

In this paper we consider a rotor for which the angular velocity is a nonconstant function of time, that is, the balls are released before the final angular velocity is attained. In particular, we focus our attention on a linearly varying angular velocity, which increases from an initial value $\omega_0$ to the final, desired angular velocity $\omega_f$, given by
\begin{equation}\label{LinearlyIncFrequency}
\omega(t) = \begin{cases} \omega_0 + (\omega_f - \omega_0)\frac{t}{T} &\mbox{if } 0 \le t < T_0, \\
\omega_f &\mbox{if } t \ge T_0,
\end{cases}
\end{equation}
where $T_0$ is the time over which the rotation speed increases. Recently there has been a growing interest in ODE systems in which a parameter initially depends on time, see for example \cite{Bart,BishopRamped,BishopRamped2,Galvanetto,Wright2,Wright,WDBG}. However, previous work has focused on systems with one-and-a-half degrees of freedom. Here we extend the work in the literature and consider a basic ADB system, which is an ODE system with higher degrees of freedom that also has vast applications in engineering and industry. Although we shall only consider a basic ADB design, the interested reader may apply the ideas presented here to modified designs.

The paper is organised as follows. In Section \ref{Sect:EOM} we derive the equations of motion for a rotating disc with an ADB attached. Whin deriving the equations the angular velocity is assumed to be time-dependent. The equations of motion are transferred to dimensionless form for the purpose of numerical implementation, which is conducted in Sections \ref{Sect:ConstantOmega} to \ref{Sect:Omega_rLessThanOmega_f}. In Sections \ref{Sect:ConstantOmega} to \ref{Sect:Omega_rLessThanOmega_f} we use numerical methods to investigate an ADB system with two balls; we will focus our attention on studying the amplitude of the transient vibrations and the percentage of phase space contained inside the basin of attraction of the damped solution, i.e. the probability of achieving balance.

In Section \ref{Sect:ConstantOmega} we study the system with constant angular velocity. This is the approach used in previous literature, which we later compare with the results in Sections \ref{Sect:Omega_rIsOmega_f} and \ref{Sect:Omega_rLessThanOmega_f}, where the angular velocity is a nonconstant function of time, given by equation \eqref{LinearlyIncFrequency}. In Section \ref{Sect:Omega_rIsOmega_f} we consider the angular velocity of the disc as an initially increasing function of time, this models the spin-up of the rotor. The balancing balls in the ADB are released once the rotor reaches the final angular velocity, $\omega_f$; this approach is similar to that in previous literature and Section \ref{Sect:ConstantOmega}, however here we are able to study the effects of rotor acceleration on the dynamics of the system. In Section \ref{Sect:Omega_rLessThanOmega_f} we consider a model in which the balls are released prior to the final angular velocity. We also briefly consider nonmonotonic variations of $\omega(t)$ and the resulting effects on the basins of attraction. Finally in Section \ref{Sect:Conclusion} we give concluding comments.

\section{Equations of motion}\label{Sect:EOM}
\begin{figure}[t]
\includegraphics[width=0.5\textwidth]{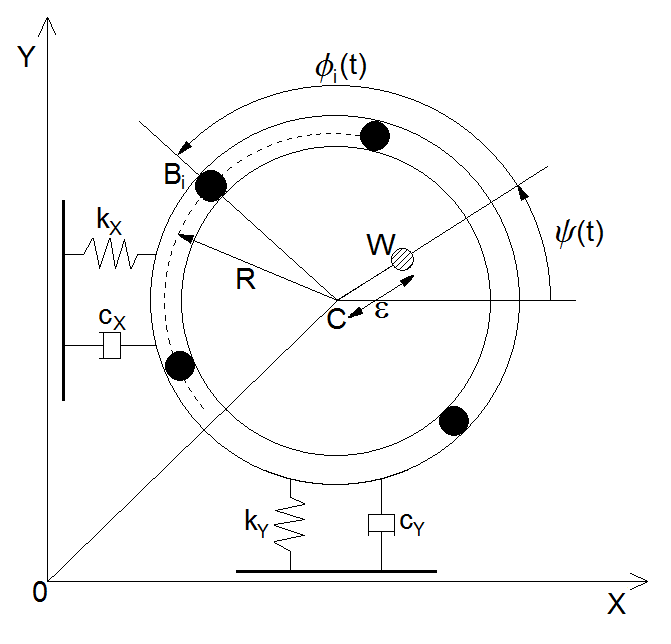}
\caption{Diagram of an ADB device attached to an eccentric disk. The coordinate $\psi(t)$ measures the angle made between the centre of mass of the disk and the horizontal axis. The coordinates $\phi_i(t)$ measure the angle between the $i$-th ball and the centre of mass of the disk. The radius of the disk is denoted by $R$ and $\varepsilon$ is the distance between the centre of mass and the centre of rotation. The parameters $c_X$ and $c_Y$ are linear damping constants which act on the rotor in the $X$ and $Y$ directions, respectively, and $k_X$, $k_Y$ are linear spring constants.}
\label{ADBDiag}
\end{figure}

The system consists of an eccentric rotating disc together with an ADB, two or more balls free to move in a race filled with a viscous fluid and positioned at a fixed distance from the centre of rotation of the disc, see Figure \ref{ADBDiag}. All motion is assumed to be confined to the two-dimensional $(X,Y)$-plane. Let $L=T-V$ be the Lagrangian with respect to this mechanical system with generalised external forces $Q=(-c_X\dot{X},-c_Y\dot{Y},\tilde{M},-D_i\dot{\phi}_i)$, where $T$ and $V$ are the kinetic energy and potential energy, respectively. The equations describing the motion of the system are the Euler-Lagrange equations
\begin{equation}\label{adbe}
\frac{\der}{\der t} \left(\frac{\partial L}{\partial \dot{q_k}}\right) - \frac{\partial L}{\partial q_k} = Q_k,
\end{equation}
where $q=(X,Y,\psi,\phi_i)$ is the generalised coordinate system. The coordinates $X(t)$, $Y(t)$ measure the displacement of the disc due
to vibrations. The coordinate $\psi(t)$ measures the angle made between the centre of
mass of the disk and the horizontal axis. The coordinates $\phi_i(t)$ measure the angle between the $i$-th ball and the
centre of mass of the disk. The parameters $c_X$ and $c_Y$ are linear damping constants which act
on the rotor in the $X$ and $Y$ directions, respectively. The value $\tilde{M}$ is the moment driving the system, and
$D_i$ is the linear drag coefficient associated with the $i$-th ball. The kinetic energy $T$ and potential energy $V$ are,
respectively, given by
\begin{equation*}
\begin{aligned}
T=&\frac{1}{2}I_z\dot{\psi}^2+\frac{1}{2}M\left[\left(\dot{X}-\varepsilon\dot{\psi}\sin\psi\right)^2
+\left(\dot{Y}+\varepsilon\dot{\psi}\cos\psi\right)^2\right]\\
&+\frac{1}{2}\sum_{i=1}^nm_i\left[\left(\dot{X}-R\left(\dot{\psi}+\dot{\phi}_i\right)\sin\left(\psi+\phi_i\right)\right)^2
+\left(\dot{Y}+R\left(\dot{\psi}+\dot{\phi}_i\right)\cos\left(\psi+\phi_i\right)\right)^2\right],
\end{aligned}
\end{equation*}
and
\begin{equation*}
V=\frac{1}{2}k_XX^2+\frac{1}{2}k_YY^2+MgY+\sum_{i=1}^nm_ig(Y+R\sin(\psi+\phi_i)),
\end{equation*}
where $M$ is the mass of the disc (without the balancing balls), $m_i$ is the mass of the $i$-th ball, $I_z$ is the moment of inertia of the rotor about the center of rotation of the disc, $R$ is the radius of the disc, $\varepsilon$ is the distance between the centre of mass and the centre of rotation of the disc, $g$ is the acceleration due to gravity, and $k_X$, $k_Y$ are linear spring constants acting on the rotor in the $X$ and $Y$ directions, respectively. The equations of motion for an ADB device with constant angular velocity have been calculated in much of the literature, see for example \cite{Chung1,Green2,Huang,Kim,Rodrigues}. We refer the reader in particular to \cite{Green2} where the equations are derived in detail. The equations of motion \eqref{EOMX}-\eqref{EOMBalls} are similar to those in the literature except for the addition of terms including $\ddot{\psi}$, which disappear when the angular velocity $\omega$ is constant. Note also that $\dot{\psi} = \omega$ only if $\omega$ is constant.
\begin{equation}\label{EOMX}
\begin{split}
& M\ddot{X} - M\varepsilon\dot{\psi}^2 \cos{\psi} - M \varepsilon \ddot{\psi}\sin{\psi}
+ \sum_{i = 1}^n m_i \left[\ddot{X} - R(\ddot{\psi} + \ddot{\phi}_i)\sin(\psi + \phi_i)\right. \\
&\qquad \qquad \qquad \qquad \qquad \qquad \qquad \qquad \qquad \left. -R(\dot{\psi} + \dot{\phi}_i)^2\cos(\psi + \phi_i)\right] + k_X X = -c_X \dot{X},
\end{split}
\end{equation}
\begin{equation}\label{EOMY}
\begin{split}
& M \ddot{Y} - M\varepsilon\dot{\psi}^2\sin{\psi} + M\varepsilon\ddot{\psi}\cos{\psi} +
\sum_{i=1}^n m_i \left[ \ddot{Y} + R(\ddot{\psi} + \ddot{\phi_i})\cos(\psi+\phi_i) \right. \\
& \qquad \qquad \qquad \qquad \qquad \qquad   \left.  - R(\dot{\psi} + \dot{\phi}_i)^2\sin(\psi + \phi_i)\right] + k_Y Y + Mg
+ \sum_{i=1}^n m_i g = -c_Y \dot{Y},
\end{split}
\end{equation}
\begin{equation}\label{EOMM}
\begin{split}
& I_z \ddot{\psi} - M\varepsilon\ddot{X}\sin{\psi} + M\varepsilon\ddot{Y}\cos{\psi} + M
\varepsilon^2\ddot{\psi} - \sum_{i=1}^n m_i\left[R\left(\ddot{X}\sin(\psi+\phi_i)\right.\right. \\
&  \qquad \qquad \qquad \qquad \qquad   \left.\left.  - \ddot{Y}\cos(\psi + \phi_i)\right) + R^2(\ddot{\psi} + \ddot{\phi}_i)\right]
+ \sum_{i=1}^n m_igR\cos(\psi + \phi_i) = \tilde{M},
\end{split}
\end{equation}
\begin{equation}\label{EOMBalls}
\begin{split}
& -m_iR\left[\ddot{X}\sin(\psi + \phi_i) - \ddot{Y}\cos(\psi + \phi_i)\right] + m_iR^2(\ddot{\psi} +
\ddot{\phi}_i)  + m_i g R \cos(\psi + \phi_i) 
= -D_i \dot{\phi}_i.
\end{split}
\end{equation}

Since the disc is driven, the angular velocity of the disc $\omega(t)$ is controlled. Therefore, the system is not governed by the torsional load and we may neglect equation \eqref{EOMM}. The remaining equations of motion may be simplified by assuming that all the balls in the balancer have
equal mass $m$ and exert equal viscous drag $D$, i.e.
\[m_i = m, \qquad D_i = D, \qquad \mbox{for } i = 1, 2, \dots, n.\]

We may non-dimensionalise the equations by setting
\[\bar{X} = \frac{X}{R}, \quad \bar{Y} = \frac{Y}{R}, \quad \bar{t} = \omega_c t,\]
and introducing the dimensionless parameters
\[\mu = \frac{m}{M}, \quad \lambda = \frac{\varepsilon}{ R}, \quad G = \frac{g}{R\omega_c^2},\]
where $\omega_c = \sqrt{k/M}$ is the critical (also called natural) angular velocity of the disk. More precisely, $\omega_c$ is the angular velocity which resonates with the springs, given the mass of the disc; it is also often referred to as the critical (or natural) frequency.
This form of $\omega_c$ assumes isotropic suspension of the rotor, namely
\[\{c,k\} = \{c_X,k_X\} = \{c_Y,k_Y\}.\]
We also introduce the following dimensionless parameters
\[\zeta = \frac{c}{\sqrt{kM}} , \qquad \beta = \frac{D}{mR^2\omega_c}.\]
The external damping ratio $\zeta$, relates to the damping induced by the springs which are attached to the outside of the ball race. The internal damping ratio $\beta$, represents the amount of drag on each ball created by the viscous fluid in the race. Note that all the parameters are assumed to be positive. This is without loss of generality since negative $\varepsilon$ implies an eccentric centre of mass located on the negative axis, which can be mapped back to the positive axis by a simple change of coordinates. The dimensionless form of equations \eqref{EOMX}, \eqref{EOMY} and \eqref{EOMBalls} is given by
\begin{equation*}\begin{split}
& (1+n\mu)\ddot{\bar{X}} + \zeta\dot{\bar{X}} + \bar{X} = \lambda \dot{\psi}^2 \cos{\psi} + \lambda \ddot{\psi} \sin{\psi} 
+\mu \sum_{i=1}^{n} \left[ (\ddot{\psi} + \ddot{\phi}_i)\sin(\psi + \phi_i) + \left(\dot{\psi} + \dot{\phi}_i\right)^2\cos(\psi + \phi_i)\right],
\end{split}
\end{equation*}
\begin{equation*}\begin{split}
& (1+n\mu)\ddot{\bar{Y}} + \zeta \dot{\bar{Y}} + \bar{Y} = \lambda \dot{\psi}^2 \sin{\psi} - \lambda \ddot{\psi} \cos{\psi} - (1 + n\mu)G 
-\mu\sum_{i=1}^{n} \Big[(\ddot{\psi} + \ddot{\phi}_i)\cos(\psi + \phi_i)\\
& \qquad \qquad \qquad \qquad \qquad \qquad \qquad \qquad \qquad \qquad \qquad \qquad \qquad \qquad \qquad - \left(\dot{\psi} + \dot{\phi}_i \right)^2\sin(\psi + \phi_i)\Big], \\
\end{split}
\end{equation*}
\begin{equation*}\begin{split}
\ddot{\phi}_i - \ddot{\bar{X}}\sin(\psi + \phi_i) + \ddot{\bar{Y}}\cos(\psi + \phi_i) + \ddot{\psi} + G\cos(\psi + \phi_i) = -\beta \dot{\phi}_i.
\end{split}
\end{equation*}
Now the dots  represent differentiation with respect to the new time $\bar{t}$. It is
easily checked that setting $\psi(\bar{t}) = \Omega \bar{t}$, with $\Omega = \omega/\omega_c$ constant, the equations of motion in \cite{Green2} are recovered. For ease we drop the `bar' notation hereafter. The system may be put into the rotating frame using the transformation
\[X = x\cos{\psi} - y\sin{\psi}, \qquad Y = x\sin{\psi} + y\cos{\psi},\]
which results in the equations
\begin{equation}\label{XYMatrixEqn}
\begin{aligned}
(1+n\mu)&\left(\begin{array}{c}
\ddot{x}\\
\ddot{y}
\end{array}\right)
+\left(\begin{array}{cc}
\zeta & -2(1+n\mu)\dot{\psi}\\
2(1+n\mu)\dot{\psi} &\zeta
\end{array}\right)
\left(\begin{array}{c}
\dot{x}\\
\dot{y}
\end{array}\right)\\
&\qquad \qquad
+\left(\begin{array}{cc}
1-(1+n\mu)\dot{\psi}^2 & -\left(\zeta\dot{\psi}+\ddot{\psi}\right)\\
\zeta\dot{\psi}+\ddot{\psi} &1-(1+n\mu)\dot{\psi}^2
\end{array}\right)
\left(\begin{array}{c}
x\\
y
\end{array}\right)=\lambda\left(
\begin{array}{c}
\dot{\psi}^2\\
-\ddot{\psi}
\end{array}
\right)\\
&\qquad \qquad
-(1+n\mu)G\left(
\begin{array}{c}
\sin\psi\\
\cos\psi
\end{array}
\right)+\mu\sum_{i=1}^n\left(
\begin{array}{cc}
\left(\dot{\psi}+\dot{\phi}_i\right)^2 & \ddot{\psi}+\ddot{\phi}_i\\
-\left(\ddot{\psi}+\ddot{\phi}_i\right) & \left(\dot{\psi}+\dot{\phi}_i\right)^2
\end{array}
\right)
\left(
\begin{array}{c}
\cos\phi_i\\
\sin\phi_i
\end{array}
\right),
\end{aligned}
\end{equation}
and
\begin{equation}\label{BallPosEqn}
\ddot{\phi}_i+\beta\dot{\phi}_i+\ddot{\psi}+G\cos(\psi+\phi_i)=\left(\ddot{x}-2\dot{\psi}\dot{y}-\dot{\psi}^2x-\ddot{\psi}y\right)\sin\phi_i
-\left(\ddot{y}+2\dot{\psi}\dot{x}-\dot{\psi}^2y+\ddot{\psi}x\right)\cos\phi_i.
\end{equation}
Taking the angular velocity as defined in \eqref{LinearlyIncFrequency}, we have
\begin{equation}\label{psiEqn}
\psi(t) = t\thinspace\Omega(t) = \begin{cases} \Omega_0 t + (\Omega_f - \Omega_0)\frac{t^2}{T_0} &\mbox{if } 0 \le t < T_0, \\
\Omega_f t &\mbox{if } t \ge T_0.
\end{cases}
\end{equation}
\section{Constant angular velocity}\label{Sect:ConstantOmega}
Throughout the remainder of the paper we restrict our attention to an ADB device with two balls,
using equations \eqref{XYMatrixEqn} and \eqref{BallPosEqn} to describe the motions of the system.
In all cases the balancing balls are released at an angular velocity $\Omega_r > \Omega_c = 1$.
As a consequence, the centrifugal forces are much greater than the gravitational forces on the
balls. Therefore we may ignore the effects of gravity, i.e. we set $G = 0$. Moreover, gravity can be neglected by assuming the rotor is held in a horizontal position.

We begin by considering an ADB device attached to a disc with constant angular velocity, which
we later compare to a disc which accelerates from a stationary position. We set the initial conditions as $x = y = \dot{x} = \dot{y} = \dot{\phi}_1 = \dot{\phi}_2 = 0$\footnote{Simply, the disc starts without any initial displacement or vibration and the balancing balls are at rest.}, and study a region $\mathcal{R}$ of phase space $(\phi_1,\phi_2)$, which is defined by $\phi_1$, $\phi_2 \in [0,2\pi)$. Throughout, the sizes of basins of attraction are given as the percentage of the region $\mathcal{R}$ which they cover. The estimates of the basin sizes are calculated using either 100 000 or 300 000 pseudo-random initial conditions chosen inside the region $\mathcal{R}$; this gives an error in the basin sizes in either the first or second decimal place with a 95\% confidence interval, see Table 1 of \cite{WDBG}.

For a balanced state to exist, the balls must be heavy enough to counteract the imbalance
of the disc. This yields the condition
\begin{equation*}
\sum_{i=1}^{2} Rm_i \ge \varepsilon M \implies \sum_{i=1}^{2} \mu_i \ge \lambda.
\end{equation*}
Note we have assumed $m_1=m_2=m$ and $\mu_1=\mu_2=\mu$.
Preliminary simulations have shown that -- provided the above condition holds -- smaller values
of $\mu$ result in lower amplitude vibrations during the transient phase. However, in industrial
applications it is likely that the amount of imbalance generated will be not known prior to using the machinery. Thus, the smaller the value of $\mu$, the more limited the device is with regard to the amount of imbalance it can rectify. Moreover, for reasons such as safety of operation, it is unlikely that machinery will operate with $\lambda$ and $\mu$ chosen such that the above equation is too close to equality. Therefore we set $\mu$ larger than necessary to counter the imbalance of the disc and fix $\lambda = 0.01$, $\mu = 0.05$ in all simulations. In \cite{Green1} the authors considered the effects of the internal and external damping ratios $\beta$ and $\zeta$, respectively. They showed that, for small fixed $\beta$, increasing $\zeta$ reduced the time-span over which transient dynamics occur. However, they also found that increasing only $\zeta$ drastically reduces the probability of achieving balance; a result of unwanted vibrations becoming asymptotically stable. Furthermore, the authors commented that taking $\zeta$ too large could lead to the loss of an attracting balanced state altogether. In preliminary simulations we found that increasing both $\beta$ and $\zeta$, the transient time-span for the system is shortened further still. Moreover, we find that also increasing $\beta$ leads to an increase in the basin of attraction corresponding to the desired balanced state, see Table \ref{ConstantOmegaTables} where $\zeta = 0.5$ and $\beta = 0.01$, 0.05, 0.1 and 0.25. Another advantage of increasing $\beta$ is that the risk of ball lag is reduced; this is discussed further in Section \ref{Sect:Omega_rLessThanOmega_f}.
\begin{table}[H]
\centering
\begin{tabular}{|ccccc|}
\hline
\multicolumn{1}{|c}{$\Omega$} & \multicolumn{1}{c}{$\beta = 0.01$} & \multicolumn{1}{c}{$\beta = 0.05$} & \multicolumn{1}{c}{$\beta = 0.1$} & \multicolumn{1}{c|}{$\beta = 0.25$} \\
\hline
1.5 & 0.00 & 0.00 & 82.85 & 94.58 \\
2.0 & 0.00 & 55.23 & 98.71 & 100.00 \\
3.0 & 10.72 & 98.32 & 100.00 & 100.00 \\
4.0 & 13.04 & 89.85 & 100.00 & 100.00 \\
5.0 & 14.56 & 95.89 & 100.00 & 100.00 \\
6.0 & 15.74 & 98.61 & 100.00 & 100.00 \\ \hline
\end{tabular}
\caption{Relative sizes of the basins of attraction corresponding to the balanced state
$x = y = 0$ for constant $\Omega$. The parameter values are fixed as $\lambda = 0.01$,
$\mu = 0.05$ and $\zeta = 0.5$, with various values for $\beta$. The sizes of the basins
of attraction are given as the percentage of the region $\mathcal{R}$ which they cover, and are calculated using 300 000 pseudo-random initial conditions.}
\label{ConstantOmegaTables}
\end{table}

It is evident from Table \ref{ConstantOmegaTables} that the ADB device works best with rotors which operate far beyond the critical angular velocity. This observation agrees with results already available in the literature, see for example \cite{Green2,Huang,Rodrigues}. It is also apparent that increasing the damping ratio $\beta$ results in an increase in the basin of attraction corresponding to the balanced state.

\begin{figure}[H]
\centering
\subfloat[]{\includegraphics[width=0.4\textwidth]{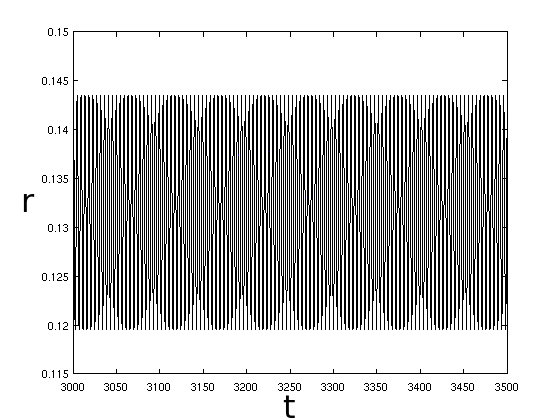}}
\subfloat[]{\includegraphics[width=0.4\textwidth]{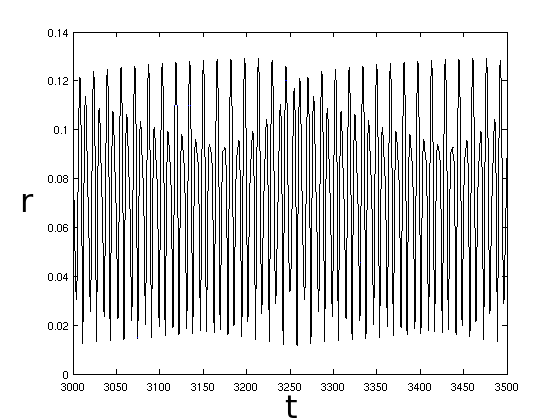}}\\
\subfloat[]{\includegraphics[width=0.4\textwidth]{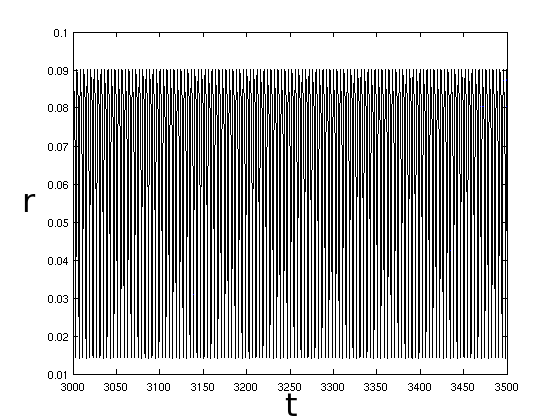}}
\subfloat[]{\includegraphics[width=0.4\textwidth]{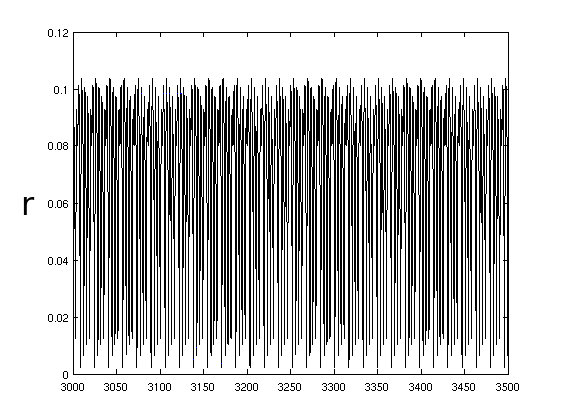}}
\caption{Plots of the radial displacement $r = \sqrt{x^2 + y^2}$ for the ``non balanced state''
attracting solutions. Figures (a) and (b) show the radial vibrations $V1$ and $V2$, respectively,
with $\Omega = 2.2$. Figures (c) and (d) show the radial vibrations $V3$ and $V4$, respectively,
with $\Omega = 4.4$. The parameter values are fixed as $\lambda = 0.01$,
$\mu = 0.05$, $\zeta = 0.5$ and $\beta = 0.05$.}
\label{V1toV4Fig}
\end{figure}

While larger $\beta$ is clearly preferable when implementing an ADB device, the system
exhibits more varied dynamics with respect to $\Omega$ when $\beta = 0.05$. As the variation is preferable when studying the effects of the acceleration of the disc on the transient and asymptotic dynamics, we shall focus our attention on the system with $\beta = 0.05$. For the chosen parameter values, four non-balanced state attractors exist, which we name $V_1$ to $V_4$. Examples of the non-balanced state attractors are given in Figure \ref{V1toV4Fig}. Of course, the exact
magnitude of the radial vibration for each $V_i$ depends on the particular value of $\Omega$.
The relative sizes of the basins of attraction for various values of $\Omega$ are given in
Table \ref{ConstantOmegaTable005}.
\begin{table}[H]
\centering
\begin{tabular}{|cccccc|}
\hline
\multicolumn{1}{|c}{$\Omega$} & \multicolumn{1}{c}{Balanced} & \multicolumn{1}{c}{$V_1$} & \multicolumn{1}{c}{$V_2$} & \multicolumn{1}{c}{$V_3$} & \multicolumn{1}{c|}{$V_4$} \\
\hline
2.0 & 55.23 & 44.77 & 00.00 & 00.00 & 00.00 \\
2.2 & 63.99 & 15.18 & 20.84 & 00.00 & 00.00 \\
2.4 & 60.54 & 03.91 & 35.54 & 00.00 & 00.00 \\
2.6 & 99.59 & 00.41 & 00.00 & 00.00 & 00.00 \\
2.8 & 99.98 & 00.02 & 00.00 & 00.00 & 00.00 \\
3.0 & 98.32 & 3$\times10^{-3}$ & 00.00 & 01.67 & 00.00 \\
3.2 & 92.55 & 3$\times10^{-6}$ & 00.00 & 07.45 & 00.00 \\
3.4 & 88.27 & 00.00 & 00.00 & 11.73 & 00.00 \\
3.6 & 88.11 & 00.00 & 00.00 & 11.89 & 00.00 \\
3.8 & 88.87 & 00.00 & 00.00 & 11.13 & 00.00 \\
4.0 & 89.85 & 00.00 & 00.00 & 10.15 & 00.00 \\
4.2 & 91.97 & 00.00 & 00.00 & 08.03 & 00.00 \\
4.4 & 93.80 & 00.00 & 00.00 & 06.19 & 6$\times10^{-5}$ \\
4.6 & 94.95 & 00.00 & 00.00 & 05.05 & 00.00 \\
4.8 & 95.42 & 00.00 & 00.00 & 04.58 & 00.00 \\
5.0 & 95.89 & 00.00 & 00.00 & 04.11 & 00.00 \\
 \hline
\end{tabular}
\caption{Relative sizes of the basins of attraction in phase space $(\phi_1,\phi_2)$
for constant $\Omega$. The parameter values are fixed at $\lambda = 0.01$, $\mu = 0.05$,
$\zeta = 0.5$ and $\beta = 0.05$. The fixed initial conditions are
$x = y = \dot{x} = \dot{y} = \dot{\phi}_1 = \dot{\phi}_2 = 0$. The sizes of the basins
are given by their percentage coverage of the region $\mathcal{R} = [0,2\pi)\times[0,2\pi)$, and are calculated using 300 000 pseudo-random initial conditions.}
\label{ConstantOmegaTable005}
\end{table}
\section{Time-dependent angular velocity with $\Omega_r \equiv \Omega_f$}\label{Sect:Omega_rIsOmega_f}
\begin{figure}[htbp]
\centering
\subfloat[]{\includegraphics[width=0.48\textwidth]{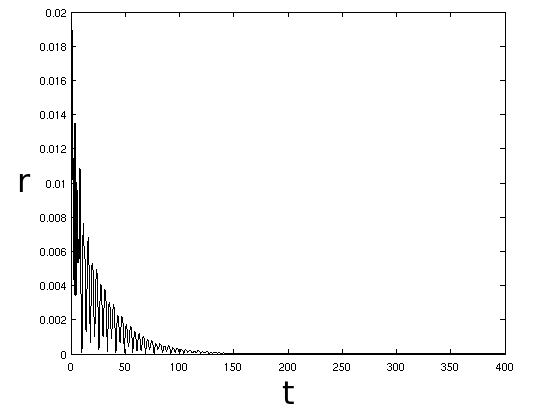}}
\subfloat[]{\includegraphics[width=0.48\textwidth]{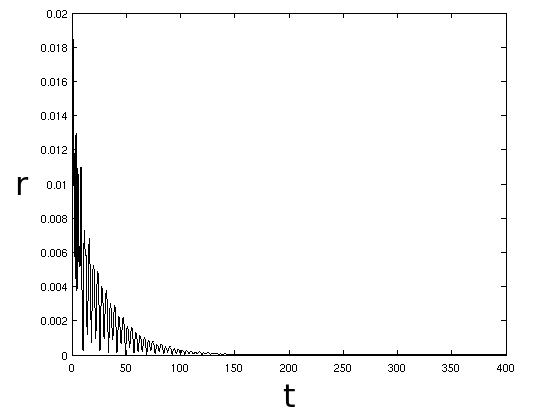}}\\
\subfloat[]{\includegraphics[width=0.48\textwidth]{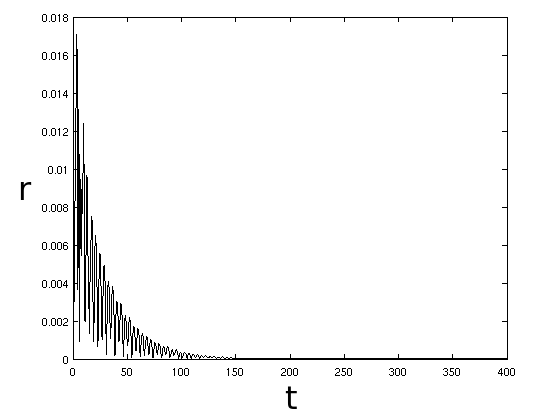}}
\subfloat[]{\includegraphics[width=0.48\textwidth]{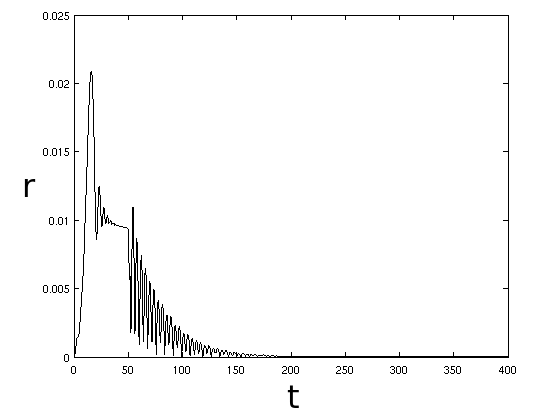}}\\
\subfloat[]{\includegraphics[width=0.48\textwidth]{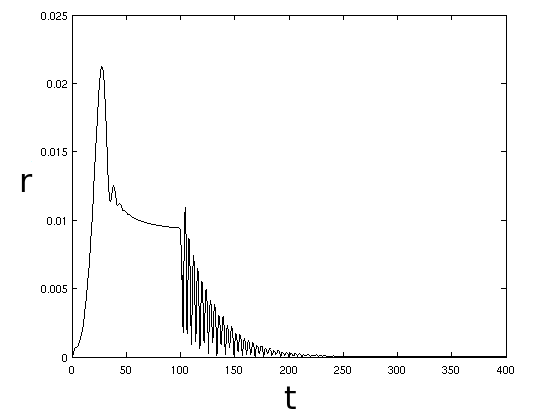}}
\subfloat[]{\includegraphics[width=0.48\textwidth]{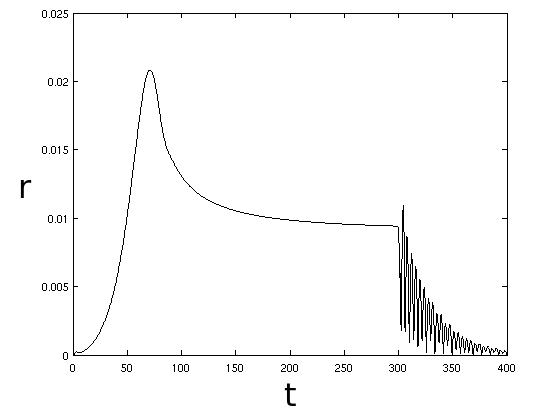}}
\caption{The radial displacement $r = \sqrt{x^2 + y^2}$ plotted against time.
The initial conditions are fixed at $x = y = \dot{x} = \dot{y} = \dot{\phi}_1 = \dot{\phi}_2 = 0$,
$\phi_1 = -\pi$, $\phi_2 = \pi$. The parameters are fixed as $\mu = 0.05$, $\zeta = 0.5$, $\lambda = 0.01$,
$\beta = 0.05$, $\Omega_0 = 0$, $\Omega_f = 2.4$. The time-spans are $T_0 = 10^{-4}$, $10^{-1}$, 5, 50, 100
and 300 in (a)-(f), respectively.}
\label{TransDynFig}
\end{figure}

We now consider an ADB device attached to a disc which accelerates in a linear manner from an initial velocity $\Omega_0$, to a final velocity $\Omega_f$, see equation \eqref{psiEqn}. To simulate a real life scenario we set $\Omega_0 = 0$, i.e. the disc accelerates from a stationary position. The time-span over which the disc accelerates is denoted by $T_0$.
Initially the balancing balls are held in position and are then released
once the disc reaches terminal velocity, that is $\Omega_r \equiv \Omega_f$ where $\Omega_r$ is the angular velocity at which the balls are released. This representation more accurately models the imbalance of the disc during the acceleration phase. Fixing $\Omega_f = 2.4$, we consider various values of $T_0$ and plot the radial displacement $r = \sqrt{x^2 + y^2}$ against time $t$, see Figure \ref{TransDynFig}. Note that the balls are located opposite each other so as not to alter the imbalance of the disc while it accelerates.

In Figure \ref{TransDynFig}(f) $T_0 = 300$: as the disc accelerates, the imbalance causes the displacement of the disc to increase. After an initial peak, the displacement settles down to an approximately constant value; this happens before the balls are released. Taking $T_0$ smaller, the dynamics happen over a shorter interval of time, see Figures \ref{TransDynFig}(d),(e) where $T_0 = 50$, 100, respectively. However, if the acceleration occurs over a very short time interval -- see Figures \ref{TransDynFig}(a)-(c) -- the system is unable to react. In this instance, the balancing balls are released whilst the radial displacement is still increasing. Hence the precise time at which the balls are released has an affect on the transient vibrations and may also alter the asymptotic solution.

The sizes of the basins of attraction for several values of $T_0$ and $\Omega_r = \Omega_f = 2.4$ are presented in Table \ref{OmegarIsOmegaf2p4}. It is clear that for $T_0$ large enough, the exact value of $T_0$ does not have any significant affect on the sizes of the basins of attraction. However, for small $T_0$ the basin sizes do differ; in this instance, with the exception of $T_0 = 1$, the basin corresponding to the balanced state is slightly larger. This is interesting as it is commonly believed that quickly accelerating up to the final angular velocity allows one to ignore the vibrations caused during the acceleration phase. In fact, the faster the disc accelerates, the further the results are from the constant $\Omega$ regime, compare the results in Table \ref{OmegarIsOmegaf2p4} with those for $\Omega = 2.4$ in Table \ref{ConstantOmegaTable005} of Section \ref{Sect:ConstantOmega}. The results of Table \ref{OmegarIsOmegaf2p4} suggest that, at least for the chosen parameter values, fast acceleration of the disc is preferable in order to maximise the chance of achieving balance. Moreover, for all values of $T_0$ shown, the sizes of the basins differ from those in Table \ref{ConstantOmegaTable005} with $\Omega = 2.4$; in particular the basin of attraction corresponding to the balanced state is approximately $10\%$ larger.
\begin{table}[htbp]
\centering
\begin{tabular}{|cccc|}
\hline
\multicolumn{1}{|c}{$T_0$} & \multicolumn{1}{c}{Balanced} & \multicolumn{1}{c}{$V_1$} & \multicolumn{1}{c|}{$V_2$} \\
\hline
$10^{-6}$ & 72.66 & 01.64 & 25.70 \\
$10^{-4}$ & 72.67 & 01.63 & 25.70 \\
$10^{-2}$ & 72.64 & 01.67 & 25.69 \\
$10^{-1}$ & 72.68 & 01.83 & 25.50 \\
1 & 68.43 & 02.41 & 29.16 \\
5 & 69.58 & 02.18 & 28.25 \\
10 & 69.55 & 02.21 & 28.24 \\
25 & 69.68 & 02.22 & 28.17 \\
50 & 69.60 & 02.25 & 28.15 \\
75 & 69.60 & 02.25 & 28.15 \\
100 & 69.60 & 02.25 & 28.15 \\
200 & 69.61 & 02.24 & 28.15 \\
300 & 69.61 & 02.24 & 28.14 \\
\hline
\end{tabular}
\caption{Sizes of the basins of attraction with $\Omega_0 = 0$, $\Omega_f = 2.4$ and various
values of $T_0$. The fixed initial conditions are
$x = y = \dot{x} = \dot{y} = \dot{\phi}_1 = \dot{\phi}_2 = 0$ and the basins are approximated
using 100 000 initial conditions in the plane $(\phi_1,\phi_2)$.}
\label{OmegarIsOmegaf2p4}
\end{table}

Simulations with $\Omega_r = \Omega_f = 5$ were also conducted. In this instance we found that the basins of attraction did not vary significantly, regardless of the value of $T_0$. In particular, the sizes of the basins presented in Table \ref{ConstantOmegaTable005}, in which $\Omega$ is constant, provide a good approximation to when $\Omega(t)$ increases from $\Omega_0 = 0$. In order to understand the underlying reason, we may think of the system as evolving over two time intervals, namely $[0,T_0)$ and $[T_0,\infty)$. During the interval $[0,T_0)$, the system is a {\it transient system}, see \cite{Galvanetto}, as the angular velocity of the disc varies with time. The evolution of initial conditions during this first time interval set up the ``initial conditions'' for the {\it steady system} in the second time interval, $[T_0,\infty)$. The ``initial conditions'' for the steady system will likely not be evenly distributed, but have formed dense clusters in phase space, depending on the attractors which exist for the {\it transient system}. The location and density of these clusters will be dependent on the precise evolution of the {\it transient system}, for more details see \cite{WDBG,Wright2}. In particular, if the basin of attraction for the balanced state is large in the {\it steady system}, it is likely to capture most of the trajectories resulting from the {\it transient system}. Therefore we may postulate that, the acceleration of the disc has more significant effects on the asymptotic behaviour of the system when the basins of attraction in the constant angular velocity regime are smaller. However, in order to rigorously apply the methods in \cite{WDBG} to understand the changes in the basins of attraction when $\Omega(t)$ varies, one must study the basins of attraction in the full phase space $(x,y,\phi_1,\phi_2,\dot{x},\dot{y},\dot{\phi}_1,\dot{\phi}_2)$\footnote{With the assumption that the balls are held in position until $t = T_0$, one may neglect the $\dot{\phi}_1$ and $\dot{\phi}_2$ directions.}. Although possible, numerically calculating the eight-dimensional basins of attraction would take a vast amount of computation time and the resulting data would be difficult to analyse. Moreover, while one could numerically implement the ideas in \cite{WDBG} to predict the asymptotic behaviour of trajectories, there is no obvious way of displaying the basins in eight dimensions, making visual analysis impossible. It is therefore not beneficial to rigorously implement the ideas of \cite{WDBG} in this case, although it can be helpful to keep them in mind.

In Table \ref{OmegarIsOmegaf} we fix $T_0 = 500$ and present results for various values of $\Omega_f$. As one may expect (given the sizes of the balanced state's basin of attraction in the constant $\Omega$ regime), for $\Omega_f \ge 2.6$ the sizes of the basin of attraction corresponding to the balanced state are similar to those in Table \ref{ConstantOmegaTable005}.

\begin{table}[htbp]
\centering
\begin{tabular}{|cccccc|}
\hline
\multicolumn{1}{|c}{$\Omega_f$} & \multicolumn{1}{c}{Balanced} & \multicolumn{1}{c}{$V_1$} & \multicolumn{1}{c}{$V_2$} & \multicolumn{1}{c}{$V_3$} & \multicolumn{1}{c|}{$V_4$} \\
\hline
2.0 & 63.88 & 36.12 & 00.00 & 00.00 & 00.00 \\
2.2 & 71.86 & 7.42 & 20.72 & 00.00 & 00.00 \\
2.4 & 69.61 & 2.24 & 28.14 & 00.00 & 00.00 \\
2.6 & 99.88 & 00.12 & 00.00 & 00.00 & 00.00 \\
2.8 & 99.98 & 00.02 & 00.00 & 00.00 & 00.00 \\
3.0 & 99.12 & 00.00 & 00.00 & 00.88 & 00.00 \\
3.2 & 94.66 & 00.00 & 00.00 & 05.34 & 00.00 \\
3.4 & 91.65 & 00.00 & 00.00 & 08.35 & 00.00 \\
3.6 & 91.38 & 00.00 & 00.00 & 08.62 & 00.00 \\
3.8 & 91.02 & 00.00 & 00.00 & 08.98 & 00.00 \\
4.0 & 90.73 & 00.00 & 00.00 & 09.27 & 00.00 \\
4.2 & 92.02 & 00.00 & 00.00 & 07.98 & 00.00 \\
4.4 & 93.89 & 00.00 & 00.00 & 06.11 & 7$\times10^{-5}$ \\
4.6 & 94.65 & 00.00 & 00.00 & 05.35 & 00.00 \\
4.8 & 95.01 & 00.00 & 00.00 & 04.99 & 00.00 \\
5.0 & 95.47 & 00.00 & 00.00 & 04.53 & 00.00 \\
 \hline
\end{tabular}
\caption{Relative sizes of the basins of attraction for $\Omega(t)$ varying from $\Omega_0 = 0$
to $\Omega_f$ over a time-span $T_0 = 500$. The balls are released at $\Omega_r = \Omega_f$. The parameter values
are set to $\lambda = 0.01$, $\mu = 0.05$ and $\zeta = 0.5$, $\beta = 0.05$. The fixed initial conditions are
$x = y = \dot{x} = \dot{y} = \dot{\phi}_1 = \dot{\phi}_2 = 0$ and the basins are approximated using 100 000
initial conditions in the plane $(\phi_1,\phi_2)$.}
\label{OmegarIsOmegaf}
\end{table}
\section{Time-dependent angular velocity with $\Omega_r < \Omega_f$}\label{Sect:Omega_rLessThanOmega_f}
We now consider releasing the balancing balls at $\Omega_r < \Omega_f$, which we directly compare with the regime studied in Section \ref{Sect:Omega_rIsOmega_f}, where $\Omega_r = \Omega_f$. That is, we compare the two scenarios:
\begin{enumerate}
\item[(i)] The angular velocity increases linearly from $\Omega(t) = \Omega_0 \equiv 0$ to
$\Omega(t) = \Omega_f$ over a time $T_0$, at which point the balls are released, that is $\Omega_r = \Omega_f$.
\item[(ii)] The angular velocity increases linearly from $\Omega(t) = \Omega_0 \equiv 0$ to
$\Omega(t) = \Omega_f$ over a time $T_0$. The balls are released at some angular velocity $\Omega_r$ with $\Omega_c < \Omega_r < \Omega_f$.
\end{enumerate}
It is clear from Table \ref{OmegarIsOmegaf} that, in scenario (i) with the chosen parameter values, when $\Omega_f \ge 2.6$ the balanced state attracts most of the phase plane $(\phi_1,\phi_2)$. This may also be seen in Table \ref{ConstantOmegaTable005}, where the acceleration phase of the disc is ignored. We therefore expect to see little or no beneficial change to the basins of attraction when considering scenario (ii) with $\Omega_f \ge 2.6$.

\begin{figure}[htb]
\includegraphics[width=0.5\textwidth]{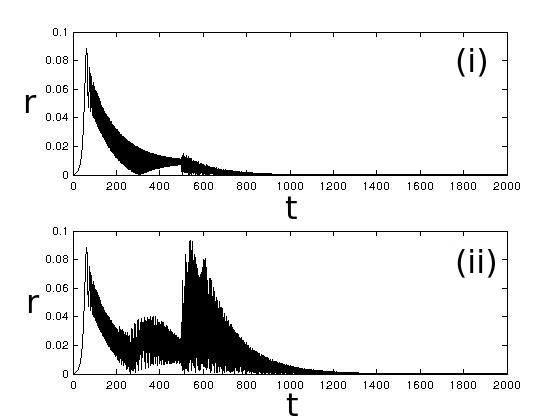}
\caption{The top plot (i) shows the vibrations when the balls are released at
$\Omega_r = \Omega_f = 5$. The bottom plot (ii) shows the vibrations for $\Omega_r = 2$. The parameters are $\lambda = 0.01$ $\mu = 0.05$, $\zeta = 0.02$,
$\beta = 0.01$ and $\Omega_0 = 0$, and we fix the initial conditions $x = y = \dot{x} = \dot{y} = \dot{\phi}_1 = \dot{\phi}_2 = 0$,
$\phi_1 = -\pi/2$, $\phi_2 = \pi/2$.}
\label{Fig4:1a}
\end{figure}

Releasing the balls before the disc reaches the final angular velocity can result in a phenomenon known as ball lag \cite{Green2}, see also \cite{Green3,Rodrigues} and the references contained therein for a more in-depth study of the phenomenon. Ball lag occurs when the balancing balls are unable to keep up with the acceleration of the disc and hence cannot balance the centre of mass at the centre of rotation; often causing vibrations far worse than the system without an ADB device. In Figure \ref{Fig4:1a} we are able to see the increased vibrations resulting from ball lag with $\zeta = 0.02$, $\beta = 0.01$; compare (i) where $\Omega_r = \Omega_f = 5$ with (ii) where the balls are released early, with $\Omega_r = 2$. However, if the internal damping $\beta$ is chosen to be sufficiently large, the effects of ball lag can be reduced; since the viscous fluid inside the race exerts enough drag force on the balls to prevent them from lagging too far behind the acceleration of the disc.

\begin{figure}[H]
\centering
\subfloat[]{\includegraphics[width=0.5\textwidth]{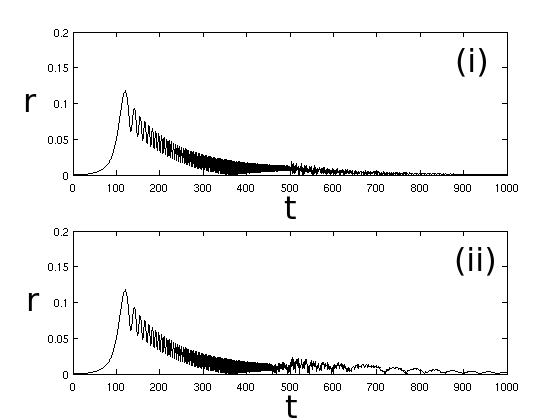}}
\subfloat[]{\includegraphics[width=0.5\textwidth]{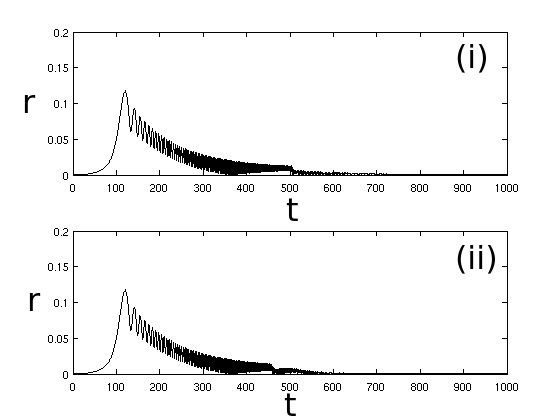}}\\
\subfloat[]{\includegraphics[width=0.5\textwidth]{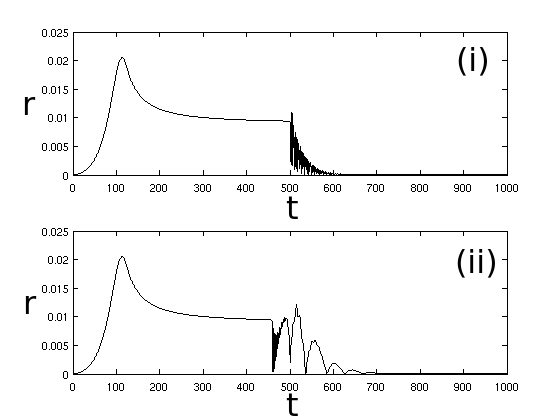}}
\subfloat[]{\includegraphics[width=0.5\textwidth]{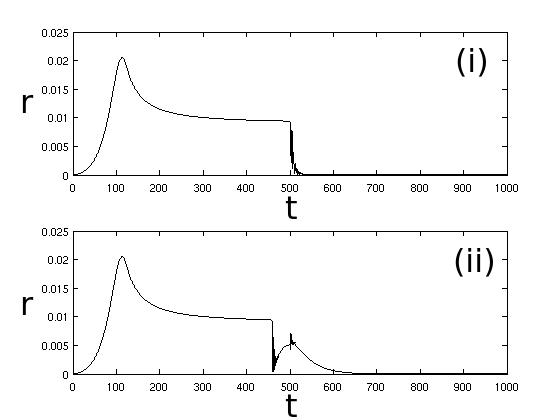}}
\caption{The top plot (i) of each figure shows the vibrations when the balls are released at $\Omega_r = \Omega_f = 2.4$.
The bottom plot (ii) shows the vibrations for $\Omega_r = 2.2$.  The parameters are $\lambda = 0.01$, $\mu = 0.05$. (a) $\zeta = 0.02$, $\beta = 0.01$.
(b) $\zeta = 0.02$, $\beta = 0.25$. (c) $\zeta = 0.5$, $\beta = 0.05$. (d) $\zeta = 0.5$, $\beta = 0.25$. We fix the initial conditions $x = y = \dot{x} = \dot{y} = \dot{\phi}_1 = \dot{\phi}_2 = 0$,
$\phi_1 = -\pi/2$, $\phi_2 = \pi/2$.}
\label{Fig4:1b}
\end{figure}

For the chosen parameters -- $\zeta = 0.5$, $\beta = 0.05$ -- ball lag does pose some risk, see Figure \ref{Fig4:1b}(c). When the balls are released at $\Omega_r < \Omega_f$, the transient vibrations whilst the disc accelerates are worse than those for the system with $\Omega_r = \Omega_f$. That said, releasing the balls early is not without benefit. For instance, when
$\Omega_r = 2.2$ and $\Omega_f = 2.4$, provided the acceleration time $T_0$ is not too small, the basin of attraction corresponding the balanced state is larger than those when $\Omega_r = \Omega_f = 2.4$. This may be seen by comparing the results in Table \ref{TabOmegar2p2Omegaf2p4} where $\Omega_r = 2.2$ with those in Tables \ref{OmegarIsOmegaf2p4} and \ref{OmegarIsOmegaf} where $\Omega_r = \Omega_f = 2.4$. Depending on the value of $T_0$ it is possible for the balanced state to attract up to 20\% more of the region $\mathcal{R}$ when the balls are released early. However, releasing the balls early does not necessarily result in preferable asymptotic dynamics. For example, in cases where $\Omega_f$ is large, the basin of attraction corresponding to the balanced state is also large when $\Omega_r = \Omega_f$, see Table \ref{OmegarIsOmegaf}. Then, releasing the balls early may result
in little or no improvement to the size of the basin of attraction, while still possibly resulting in worse transient dynamics. Furthermore, releasing the balls early can reduce the size of the desired basin of attraction. For example, Table \ref{TabOmegar23Omegaf5} shows
results for the system with $\Omega_r = 2$, 3 and $\Omega_f = 5$; in all cases the safe basin is smaller as a result of releasing the balls early, compared with $\Omega_r = \Omega_f = 5$, see Table \ref{OmegarIsOmegaf}.

\begin{table}[H]
\centering
\begin{tabular}{|cccc|}
\hline
\multicolumn{1}{|c}{$T_0$} & \multicolumn{1}{c}{Balanced} & \multicolumn{1}{c}{$V_1$} & \multicolumn{1}{c|}{$V_2$} \\
\hline
100 & 48.34 & 05.90 & 45.76 \\
200 & 55.71 & 05.55 & 38.74 \\
300 & 72.17 & 03.99 & 23.84 \\
400 & 86.61 & 03.51 & 09.89 \\
500 & 89.40 & 03.36 & 07.23 \\
600 & 90.59 & 02.94 & 06.46 \\
700 & 91.83 & 02.83 & 05.80 \\
800 & 91.93 & 02.41 & 05.65 \\
\hline
\end{tabular}
\caption{Relative sizes of the basin of attraction with $\Omega_0 = 0$, $\Omega_r = 2.2$, $\Omega_f = 2.4$ and various choices of $T_0$. The parameters are fixed as $\mu = 0.05$, $\zeta = 0.5$, $\lambda = 0.01$ and $\beta = 0.05$. The relative sizes of the basins were calculated using 100 000 pseudo-random initial conditions inside the region $\mathcal{R}$.}
\label{TabOmegar2p2Omegaf2p4}
\end{table}
\begin{table}[H]
\centering
\begin{tabular}{|ccc|}
\hline
\multicolumn{1}{|c}{$T_0$} & \multicolumn{1}{c}{$\Omega_r = 2$} & \multicolumn{1}{c|}{$\Omega_r = 3$}\\
\hline
100 & 14.92 & 49.34 \\
150 & 50.13 & 77.86 \\
200 & 68.63 & 84.44 \\
300 & 81.18 & 89.62 \\
400 & 84.31 & 93.03 \\
500 & 85.66 & 94.27 \\
1000 & 88.16 & 96.05 \\ \hline
\end{tabular}
\caption{Relative sizes of the basin of attraction corresponding to the balanced state $x = y = 0$. The parameters are set to $\mu = 0.05$, $\zeta = 0.5$, $\lambda = 0.01$ and $\beta = 0.05$. The results shown are for $\Omega_r = 2$, 3, with $\Omega_0 = 0$ and $\Omega_f = 5$. The relative sizes of the basins were calculated using 300 000 pseudo-random initial conditions inside the region $\mathcal{R}$.}
\label{TabOmegar23Omegaf5}
\end{table}
Let us now briefly consider a nonmonotonic model for the angular velocity. It may be seen from Table \ref{OmegarIsOmegaf} that, for the chosen parameters, the balanced state has a much larger basin of attraction when $\Omega_f = 3$ than for $\Omega_f = 2$. It may be the case that one wishes to set $\Omega_f = 2$, without sacrificing such a large portion of the basin of attraction. This can be achieved by increasing the angular velocity to $\Omega(t) = \Omega_r = 3$, maintaining the higher rotation speed for a short period of time then decelerating the angular velocity of the disc to the desired rotation speed $\Omega_f = 2$. This model of the angular velocity may be written as
\begin{equation}\label{NonMonOmegaEqn}
\Omega(t) = \begin{cases} \Omega_0 + (\Omega_r - \Omega_0)\frac{t}{T_0} &\mbox{if } 0 \le t < T_0, \\
\Omega_r &\mbox{if } T_0 \le t < T_1, \\
\Omega_r + (\Omega_f - \Omega_r)\frac{t-T_1}{T_2 - T_1} &\mbox{if } T_1 \le t < T_2, \\
\Omega_f &\mbox{if } t \ge T_2 ,
\end{cases}
\end{equation}
so that the system has four distinct time-intervals $[0,T_0)$, $[T_0,T_1)$, $[T_1,T_2)$ and $[T_2,\infty)$. If the disc attains balance at $\Omega(t) = \Omega_r = 3$, then provided the system does not suffer with ball lag, or indeed the balls move too fast when the disc decelerates, the disc should remain balanced when $\Omega(t) = \Omega_f = 2$, see \cite{Wright2,Wright,WBDG} for further details. By suitably manipulating the angular velocity and choosing the values of $T_0$, $T_1$ and $T_2$, one is able to obtain a large basin of attraction for the balanced state with $\Omega_f = 2$, see Table \ref{NonmonotonicOmegaTable}. In fact, on comparison with Table \ref{OmegarIsOmegaf} we find approximately a 30\% increase in the basin's size, and a 40\% increase compared with the predictions of Table \ref{ConstantOmegaTable005} where $\Omega$ is constant. 
\begin{table}[H]
\centering
\begin{tabular}{|cccc|}
\hline
\multicolumn{1}{|c}{$T_0$} & \multicolumn{1}{c}{$T_1$} & \multicolumn{1}{c}{$T_2$} & \multicolumn{1}{c|}{Basin size} \\
\hline
50 & 400 & 500 & 98.43 \\
100 & 300 & 500 & 97.19 \\
100 & 400 & 500 & 97.64 \\
200 & 400 & 500 & 93.64 \\
\hline
\end{tabular}
\caption{Sizes of the basin of attraction corresponding to the balanced state $x = y = 0$. The parameters are fixed as $\mu = 0.05$, $\zeta = 0.5$, $\lambda = 0.01$ and $\beta = 0.05$. The angular velocity is modelled by equation \eqref{NonMonOmegaEqn} with $\Omega_0 = 0$, $\Omega_r = 3$ and $\Omega_f = 2$. The relative sizes of the basins were calculated using 300 000 pseudo-random initial conditions inside the region $\mathcal{R}$.}
\label{NonmonotonicOmegaTable}
\end{table}
\section{Concluding remarks}\label{Sect:Conclusion}
We have compared the dynamics for the system of an ADB attached to a rotating disc with both constant and nonconstant angular velocities. For the parameter values investigated we numerically showed that provided the final angular velocity is high enough, one may ignore the spin-up phase of the rotor and still obtain reasonable estimates to the basin sizes. However, we also noted that this is likely due to the large size of the basin of attraction corresponding to the balanced solution; if a system is such that the balanced solution has a small basin of attraction for high angular velocities it may not be possible to ignore the spin-up phase of the rotor. 

In cases where the angular velocity is slower, and hence the basin of attraction corresponding to damped vibrations is smaller, the initial acceleration of the disc is important when studying the asymptotic dynamics. Furthermore, accelerating quickly up to the final angular velocity does not allow the scenario of constant angular velocity to serve as a good approximation to the asymptotic dynamics.

Releasing the balancing balls early can result in a higher chance of achieving balance when the system operates at lower velocities, i.e. when the desired basin is smaller. However, it can also result in worse transient dynamics. At the end of Section \ref{Sect:Omega_rLessThanOmega_f} we briefly considered nonmonotonic variations of the angular velocity and showed that, by suitable manipulation, one can significantly increase the size of the desired basin of attraction, corresponding to damped vibrations.

In addition we have shown that increasing the internal damping in conjunction with the external damping results in shorter transient times and an increase in the size of the basin of attraction corresponding to the desired solution. This is also beneficial as it allows one to minimise the risk of undesired vibrations resulting from ball lag. This may be coupled with releasing the balls early to obtain short transient time-spans and a large basin of attraction corresponding to damped vibrations. In mechanical systems it is possible to control the internal damping by increasing/decreasing the viscosity of the liquid in the ball race. Therefore, in industrial applications, it should not be difficult to implement the ideas presented here.

This work constitutes the first steps to investigating an ADB device attached to a disc with varying angular velocity. In the process, we have extended the work already available in the literature which is concerned with time-dependent parameters in ODE systems with one-and-a-half degrees of freedom, to systems with higher degrees of freedom. It would be interesting to study the system further, in particular, one could study the basins of attraction in the plane $(\lambda, \phi_1)$ with $\phi_2$ fixed as $\phi_1  + \pi$. It would also be interesting to consider the parameter plane $(\lambda, \mu)$ with $\phi_1$, $\phi_2$ fixed. Such studies would allow engineers to predict the amount of eccentricity which could be balanced, given a particular ADB set up.
\section*{Acknowledgements}
 LP is partially supported by JSPS Grant-in-Aid for Scientific Research (No. 16KT0024), the MEXT `Top Global University Project', Waseda University Grant for Special Research Projects (No. 2019C-179, No. 2019E-036) and Waseda University Grant Program for Promotion of International Joint Research.


\end{document}